\documentclass[12pt,3p,preprint,sort&compress]{elsarticle}
\pdfoutput=1
\usepackage[colorlinks=true]{hyperref}

\usepackage{amsmath}
\usepackage{amssymb}
\usepackage{mathtools}

\renewcommand{\l}{\left(}
\renewcommand{\r}{\right)}

\newcommand{\be}{\begin{equation}}
\newcommand{\ee}{\end{equation}}
\newcommand{\ba}{\begin{align}}
\newcommand{\ea}{\end{align}}
\newcommand{\bg}{\begin{gather}}
\newcommand{\eg}{\end{gather}}
\newcommand{\bseq}{\begin{subequations}}
\newcommand{\eseq}{\end{subequations}}

\begin{document}
\begin{flushright}
	INR-TH-2020-035
\end{flushright}

\title{Refining constraints from Borexino measurements on a light $Z'$-boson coupled to $L_\mu$-$L_\tau$ current} 
\author[inr]{Sergei Gninenko}
\ead{Sergei.Gninenko@cern.ch}
\author[inr,mpti]{Dmitry Gorbunov}
\ead{gorby@ms2.inr.ac.ru}
\address[inr]{Institute for Nuclear Research of Russian Academy of Sciences, 117312 Moscow, Russia}
\address[mpti]{Moscow Institute of Physics and Technology, 141700 Dolgoprudny, Russia}
\begin{abstract}
The recent confirmation by FNAL of the $(g-2)_\mu$ muon anomaly gives
strong evidence for the possible existence of new physics beyond the
Standard Model in the muon sector. Thus it is worthy to revisit the
existing experimental constraints on models suggesting theoretically
consistent explanations of the anomaly. In this work, we point out
that accounting for the loss of coherence between the wave packets
(mass states) of solar neutrinos is important for setting limits on
any model with new flavor-sensitive couplings in the neutrino sector.
By taking into account this effect and considering more accurately the
experimental constraints from the BOREXINO measurement of the ${}^7$Be
solar neutrino interaction rate we corrected the limits previously
placed on the coupling of the light $Z'$ to $L_\mu$-$L_\tau$ current
in the parameter space relevant to the muon $(g-2)_\mu$.
 \end{abstract}

\date{}

\maketitle

{\bf 1.}  A longstanding problem in particle physics naturally
stimulates theorists to invent solutions beyond the Standard Model of
particle physics (SM). Discrepancies between measurements and SM
predictions can indeed be because of the incompleteness of the SM
(though, overlooked experimental issues and problems with theoretical
calculations are not excluded generically) and indicate explicit
effects of new physics. A well known example is a tension in the muon
anomalous magnetic moment, $a_\mu = (g-2)_\mu/2$, see, e.g,
\cite{g-2_Review}, inspiring this activity for two decades. The recent
measurement of the $a_\mu$ value by the E989 experiment at Fermilab
shows a discrepancy with respect to the theoretical prediction of the
SM \cite{fnal} which when combined with the previous
Brookhaven results \cite{bnl} leads to a $4.2 \sigma$ tension of
$\Delta a_\mu = 251(59)\times 10^{-11}$.

One of the possible solutions of $(g-2)_\mu$ puzzle is a new massive
vector boson $Z'$ coupled to non-anomalous difference of muon and tau
lepton currents, $L_\mu$-$L_\tau$ \cite{foot1}-\cite{vecmuon3},
\begin{equation}
  \label{Z-flavor}
        {\cal L}=g_{Z'}Z_\rho \l L_\mu^\rho - L_\tau^\rho\r \equiv
      g_{Z'}Z_\rho \l \bar \tau \gamma^\rho\tau +
      \bar\nu_\tau\gamma^\rho\nu_\tau
      - \bar \mu \gamma^\rho\mu -
      \bar\nu_\mu\gamma^\rho\nu_\mu\r\,.  
  \end{equation}
The
electrophobic nature of $Z'$ prevents it from contributing to the
similar $(g-2)_e$ observable of electron, which otherwise would clash
with measurements. While limited by physical constraints from many
experiments in particle physics and astrophysics revealing no
anomalies, $Z'$ boson can play a role of messenger of the dark matter
sector, which makes the model interesting in cosmology and deserving
to be tested in a wider range of model parameters
\cite{gkm}-\cite{Araki:2015mya}. In particular, the promising for $(g-2)_\mu$ part of model parameter
space can be fully explored by the  NA64 experiment operating at the CERN SPS \cite{gkm, na64proposal, sgnk}.  

{\bf 2.}    One important limit \cite{kamada1} comes from
measurements of ${}^7$Be solar neutrino interaction rate inside the
BOREXINO detector\,\cite{Bellini:2011rx}: virtual intermediate $Z'$
boson may contribute to $\nu_\mu - e$ and $\nu_\tau - e$ scatterings off
electrons in the detector volume. The $Z'$ coupling to fermions of the first
generation emerge due to the $\gamma - Z'$ kinetic mixing, which is generated through the loop involving the muon and tau leptons \cite{sgnk,kamada1,Araki:2015mya}.  Consistency of the measured rate with SM predictions has allowed for imposing limits from above on the magnitude of $Z'$ boson coupling $g_{Z'}$ to $L_\mu$-$L_\tau$. 

However, in doing it, the authors, e.g.,  of Ref.\,\cite{kamada1}
have  implicitly assumed that  ${}^7$Be solar neutrinos, being originally pure electron neutrinos $\nu_e$, oscillate on their way to Earth and arrive there as equipartition combination of pure flavor states $\nu_e$, $\nu_\mu$ and $\nu_\tau$, so that each of the neutrino flavor \emph{interacts coherently} with matter of BOREXINO detector. The interaction rate then is given by  
\begin{equation}
    \label{4*}
    \Gamma \propto F_{\nu_\mu}+F_{\nu_\tau}=\frac{2}{3}F_{{}^7\text{Be}}\,, 
\end{equation}
where $F_{\nu_\alpha}$ are flavor fluxes and $F_{{}^7\text{Be}}$ is electron neutrino flux from ${}^7$Be to be measured in absence of neutrino oscillations. 

In fact, that is not the case, since the neutrino wave packets lose the coherence on their way to Earth \cite{Tanabashi:2018oca} and arrive as pure mass states $\nu_i$, $i=1,2,3$ with fluxes 
\begin{equation}
  \label{fluxes}
F_i=\left|U_{ei}\right|^2 F_{{}^7\text{Be}}\,,
\end{equation}
where $U_{\alpha i}$ are elements of the PMNS matrix\,\cite{Tanabashi:2018oca}. 
Each of the mass state is a known combination of flavor states,
$\nu_i=\sum_iU_{\alpha i}\nu_\alpha$, which hence couple coherently
when exchanging $Z'$ with matter. Since $\nu_e$ does not couple to
$Z'$ boson (except through the mentioned mixing with $Z$ boson which
is small) and $\nu_\mu$ and $\nu_\tau$ coupling constants are equal in
magnitude but of opposite signs, the interaction of the neutrino mass
states with $Z'$ following from \eqref{Z-flavor} upon rotation of the
neutrino states reads
\begin{equation}
  \label{Z-mass}
        {\cal L}=
      g_{Z'}Z_\rho 
      \bar\nu_i\gamma^\rho\nu_j\l U^*_{\tau i} U_{\tau j} -U^*_{\mu i} U_{\mu j} \r .
  \end{equation}
Hence, 
the expected impact of $Z'$ on the ${}^7$Be solar neutrino interaction
rate in BOREXINO should be summed over all the three initial mass
states and all the three final neutrino mass states,
which reveals
\begin{equation}
  \label{correct-rate}
\begin{split}
\Gamma &\propto \sum_i  F_i\times \sum_j\left|U^*_{\mu
  i}U_{\mu j} - U^*_{\tau i}U_{\tau j}\right|^2
=F_{{}^7\text{Be}}\times \sum_i  \left|U_{ei}\right|^2\times \sum_j\left|U^*_{\mu
  i}U_{\mu j} - U^*_{\tau i}U_{\tau j}\right|^2 
\\&=F_{{}^7\text{Be}}\times \sum_i  \left|U_{ei}\right|^2\times \l
\left|U_{\mu i}\right|^2 + \left|U_{\tau i}\right|^2\r =
F_{{}^7\text{Be}}\times \sum_i  \left|U_{ei}\right|^2\times \l 1- \left|U_{ei}\right|^2\r  
\\&=F_{{}^7\text{Be}}\times \l 1-\sum_i  \left|U_{ei}\right|^4\r,
\end{split}
\end{equation}
where we simplify the formula by making use of the
unitarity of the PMNS matrix. Instead of \eqref{4*}, it gives numerically 
\[
\approx 0.45\times F_{{}^7\text{Be}} 
\]
for the central values of the neutrino mixing parameters taken from
\cite{Tanabashi:2018oca}.

 One observes, that the combination of the PMNS matrix elements
  emerging in the results \eqref{correct-rate} is equal to the electron survival
  probability $P_{ee}=\sum _i
  \left|U_{ei}\right|^4$ in case of pure vacuum oscillations. Recall, that ${}^7$Be solar neutrino flux
  is expected to be a bit suppressed by scattering off 
  electrons inside the Sun: neutrino mixing angles in matter is
  somewhat different for ${}^7$Be neutrino energy as compared to pure
  vacuum case used for the numerical estimate above. 
  However the theoretical expectation and
  observations by BOREXINO\,\cite{Bellini:2011rx} are both consistent
  within the errors with pure vacuum case, which we adopt here to be
  conservative in placing limits on the new physics.

Note in passing that the same formulas \eqref{correct-rate} can be
obtained by considering the neutrino interaction in the flavor
basis. There the scattering of muon and tau neutrinos from a
particular wave packet (pure mass state) must be sum up incoherently,
which gives
\[
\Gamma \propto \sum_i  F_i\times
\l\left|U_{\mu i}\right|^2 + \left|U_{\tau i}\right|^2\r
\]
coinciding with \eqref{correct-rate} for the fluxes \eqref{fluxes}. 
Consequently, the limits of $Z'$ boson couplings $g_{Z'}$ from
Ref.\cite{kamada1} must be weakened by a factor $\approx
(0.66/0.45)^{1/4}\approx 1.11$. 

{\bf 3.}   The BOREXINO limit\,\cite{Bellini:2011rx} has been applied in
Ref.\,\cite{kamada} in a different way to infer constraints on the $L_\mu
- L_\tau$  model parameters. The authors took the central value of the
measured by BOREXINO survival probability for ${}^7$Be solar
neutrinos, $P_{ee}=0.51$ (so the $\nu_\mu$ 
and  $\nu_\tau$ flux is $0.49\times F_{{}^7\text{Be}}$)
and stated that the new physics contribution should not exceed 8\% \cite{kamada}. 
This estimate should be also corrected. First, for the best fit values
of the neutrino sector parameters and Standard Solar Model (SSM) the 
theoretical evaluation of the electron survival probability is somewhat
higher than the measured number, resulting in the  the $\nu_\mu$ 
and  $\nu_\tau$ flux of $0.47\times F_{{}^7\text{Be}}$, see Fig. 2 in Ref.\,\cite{Bellini:2011rx}, 
which is still consistent with pure vacuum oscillations as mentioned above.
Second, the BOREXINO estimate for
the survival $\nu_e$  probability including both theoretical SSM and
experimental uncertainties, $P_{ee} = 0.51\pm 0.07$,  has the total
error of 14\%  \cite{Bellini:2011rx}. This bigger error accounts
  also for the uncertainties in the flux prediction associated with
  the Standard Solar Model, which apparently have been ignored in Ref.\,\cite{kamada}. 
 Hence, to place a limit on the possible $L_\mu - L_\tau$ $Z'$  contribution, one should adopt that is should not exceed  23\% at 90\% C.L..
 Consequently, the estimate in Ref.\,\cite{kamada} on $g'$ coupling has to be corrected by a factor
\begin{equation}   
\l\frac{0.23}{0.08}\times \frac{0.49}{0.47}\r^{1/4}=1.32\,. 
\label{corr-factor}
\end{equation}
 \begin{figure}[htb!]
\begin{center}
\includegraphics[width=0.6\textwidth,height=0.5\textwidth]{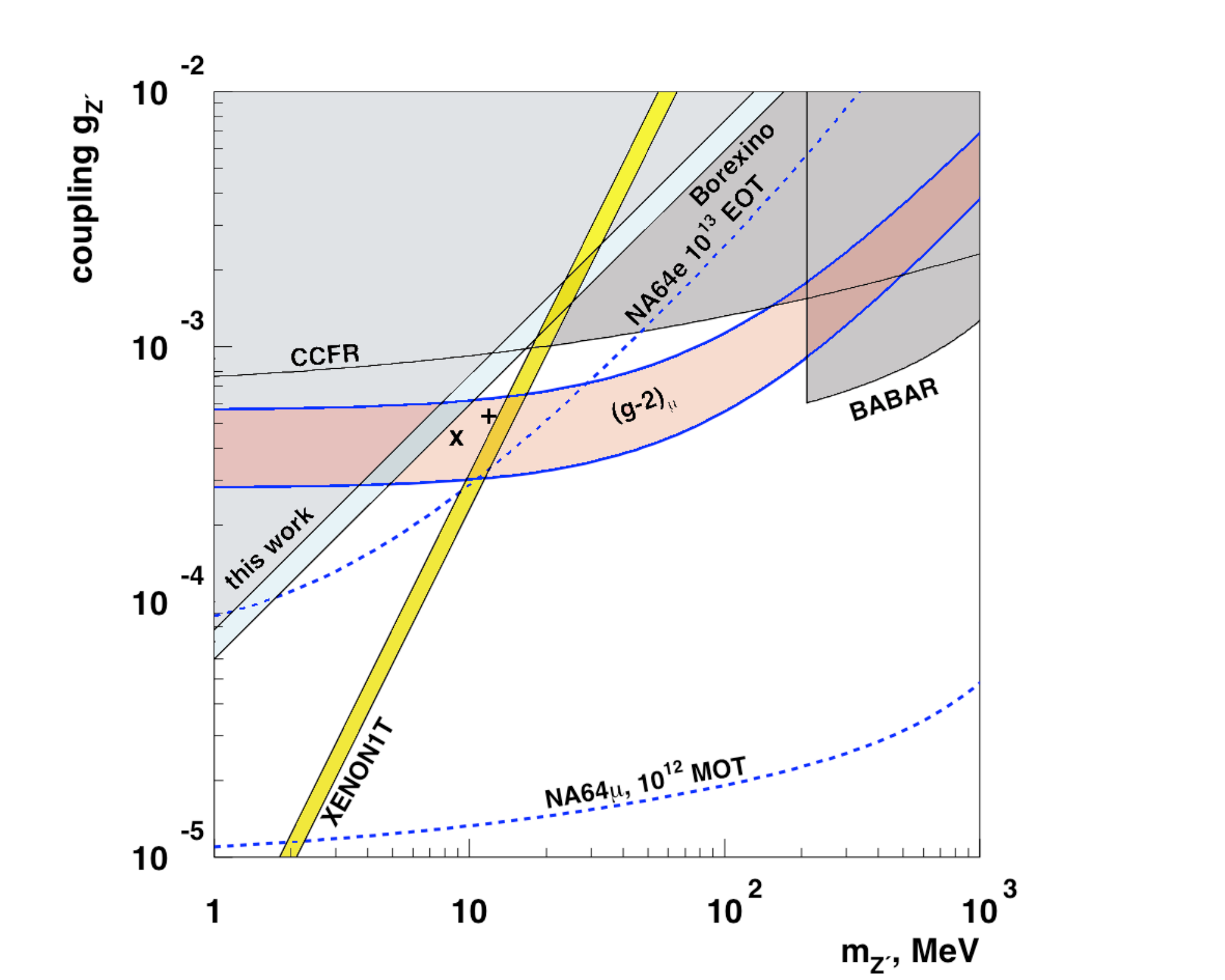}
\caption {Shown are the 90\% C.L. exclusion regions (shaded areas above the curves)  from the analysis of  BOREXINO data performed in \cite{kamada} and in this work (light grey area)  in the ($m_{Z'}, g_{Z'}$)  plane
together with the favourable $(g-2)_\mu$ parameter space in the $L_\mu - L_\tau$ model. 
The 90\% C.L. expected exclusion regions(dashed curves) from the measurements with the elections with  $\simeq 10^{13}$  EOT\cite{sgnk, na64prl}  and  muons with   $\simeq 10^{12}$ MOT  \cite{gkm,na64proposal}  in the NA64 experiment at CERN, as well as 
 constraints from the CCFR   \cite{ccfr}, and BABAR \cite{BABAR1} experiments, and the region of parameters 
 explaining the XENON1T excess \cite{x1t} are also shown.  Two symbols {\bf +} and {\bf x} 
  indicate parameters $(m_{Z'}, g_{Z'})$ = (11 MeV, $5\times 10^{-4}$) and (9 MeV, $4\times 10^{-4})$, respectively, 
 which are used to explain the characteristic features of cosmic neutrino spectrum reported by the IceCube Collaboration, see Ref.\cite{Araki:2015mya} for details. 
 \label{fig:excl}}
\end{center}
\end{figure} 
This results in a new  90\% C.L. exclusion region shown  in the ($m_{Z'}, g_{Z'}$)  plane in Fig.\,\ref{fig:excl}, 
 together with the  BOREXINO  limit from Ref.\cite{kamada} and the favourable $(g-2)_\mu$ parameter space of  the $L_\mu - L_\tau$ model. 
The existing constraints from the CCFR   \cite{ccfr} and BABAR \cite{BABAR1} experiments,  as well as expected  limits from the measurements with the election beam for  $\simeq 4\cdot 10^{13}$ electrons on target (EOT)  \cite{na64prl}  and the muon beam for  $\simeq 10^{12}$  muons on target (MOT)  \cite{gkm}  with the NA64 experiment at CERN  are also shown. 
Interestingly, in a specifically extended model there is a region in the ($m_{Z'}, g_{Z'}$) plane which could simultaneously explain
 the $(g-2)_\mu$ and an excess of low energy electrons recently observed by the XENON1T experiment \cite{x1t}.
It is worth mentioning that particular variants of the model with light $Z'$-bosons are generically constrained from cosmology, e.g. due to possible impact of light $Z'$-bosons on the Big Bang Nucleosynthesis and Comic Microwave Background anisotropies \cite{kamada, kamada1, hooper}.
Nevertheless,  this observation makes searches for $Z'$ with NA64 quite interesting, as the  experiment  can probe 
 the parameter space shown in Fig.\ref{fig:excl} and  provide
 important complementary results in the near future \cite{na64proposal}. As shown in Fig.\,\ref{fig:excl}
 with about $\sim 10^{13}$ EOT  NA64e will probe  the area of parameters explaining both, the $(g-2)_\mu$ anomaly and the XENON1t excess. Interestingly, it will also probe both   
  reference points of the model suggested in Ref.\,\cite{Araki:2015mya} for explaining characteristic features of cosmic neutrino spectrum 
  observed by  IceCube, which are also shown in  Fig.\,\ref{fig:excl}.
\par  To conclude, from the theoretical point of view we expect  the effect of losing the coherence is  particularly
important for the phenomenology of any model of new physics with
non-universal interactions with neutrino flavors: e.g. in models with neutrino
dipole moments different in values for electron, muon and tau
neutrinos. While from the experiment view point, given the significance of the (g-2)$_\mu$ anomaly for a possible discovery 
of new physics, correction \eqref{corr-factor} enlarges the  ($m_{Z'}, g_{Z'}$) parameter space for the  $L_\mu - L_\tau$ $Z'$ explanation 
of the muon $(g-2)_\mu$ anomaly, and at the same time increases motivation  for  searching the $Z'$ in NA64e and  NA64$_\mu$ experiments at CERN \cite{gkm, na64proposal, sgnk}  and 
M$^3$ \cite{krnjaic} at Fermilab, in particular in the parameter space,  which could explain the $(g-2)_\mu$
anomaly and the XENON1T excess simultaneously. This makes these searches quite exciting.  Despite of the
fact that BOREXINO bounds have already probed indirectly the scenario
under consideration, it would be of great importance to perform in
these experiments independent direct searches for the $Z'$-boson in the mass range $m_{Z'} \lesssim 10$ MeV.
 An estimate shows that, e.g.,  the NA64e constraints on the  couplings $g_{Z'}$   in this  mass range, which can be obtained from the analysis of data collected for  the invisible decays of dark photons \cite{sgnk, na64prl},  are comparable with those presented above. 
\vskip 0.3cm
We thank Mikhail Shaposhnikov for useful discussions. 
The estimate of the interaction rate at BOREXINO was supported 
by the RSF Grant No. 17-12-01547. The work on NA64  was supported in part by the Ministry of Science and Higher Education of Russia, the Agreement  No 05.613.21.0098 ID No RFMEFI61320X0098 on July 23,  2020.


\end{document}